\documentclass[twocolumn,showpacs,showkeys,preprintnumbers,superscriptaddress,amsmath,floatfix,amssymb,secnumarabic,nofootinbib]{revtex4}

\usepackage[colorlinks=true]{hyperref}
\usepackage{graphicx}
\usepackage{braket}
\usepackage{amsmath}
\usepackage{epsfig}
\usepackage{float}
\usepackage{nicefrac}
\usepackage{xcolor}
\usepackage{color}

\usepackage{bm}

\def\({\left(}
\def\){\right)}
\def\[{\left[}
\def\]{\right]}

\newcommand{\vev}[1]{ \langle \, #1 \, \rangle }
\newcommand{\diag}[1]{ {\rm diag} \, \left( #1 \right) }

\newcommand{\beq} {\begin{eqnarray}}
\newcommand{\eeq} {\end{eqnarray}}

\newcommand{\comment}[1]{}

\begin{document}
\sloppy

\title{Taming the sign problem of the finite density Hubbard model via Lefschetz thimbles}

\author{Maksim~Ulybyshev}
\email{Maksim.Ulybyshev@physik.uni-wuerzburg.de}
\affiliation{Institute of Theoretical Physics, Julius-Maximilians-Universit\"at,
   97074 W\"urzburg, Germany}

\author{Christopher Winterowd}
\email{c.r.winterowd@kent.ac.uk}
\affiliation{University of Kent, School of Physical Sciences, Canterbury CT2 7NH, UK}

\author{Savvas Zafeiropoulos}
\email{s.zafeiropoulos@thphys.uni-heidelberg.de }
\affiliation{Institute for Theoretical Physics, Heidelberg University, Philosophenweg 12, 69120 Heidelberg, Germany}

\begin{abstract}
We study the sign problem in the Hubbard model on the hexagonal lattice away from half-filling using the Lefschetz thimbles method. We identify the saddle points, reduce their amount, and perform quantum Monte Carlo (QMC) simulations using the holomorphic gradient flow to deform the integration contour in complex space. Finally, the results are compared against exact diagonalization (ED).  We show that the sign problem can be substantially weakened, even in the regime with low temperature and large chemical potential, where standard QMC techniques exhibit an exponential decay of the average sign.
\end{abstract}
\pacs{11.15.Ha, 02.70.Ss, 71.10.Fd}
\keywords{Hubbard model, sign problem, Lefschetz thimbles}

\maketitle

{\it Introduction.}
Monte Carlo calculation of the Feynman path integral is one of the most reliable methods for non-perturbative study of the physics of strongly coupled systems in a fully \textit{ab-initio} manner. The conventional lore states that after performing a Wick rotation from Minkowski to Euclidean space, the Euclidean action of the system is real and thus the integral is no longer oscillatory. Thus, its evaluation can proceed in standard fashion using importance sampling. Unfortunately, there are many interesting systems where the Euclidean action is complex. In high energy physics, we have the example of quantum chromodynamics at finite baryon density~\cite{Philipsen:2010gj,Aarts:2015tyj}. The sign problem also impedes \textit{ab-initio} studies of condensed matter and atomic systems. An example of the latter is the unitary Fermi gas (UFG), a system materialized frequently in the laboratory~\cite{Inguscio:2007cma,Schafer:2009dj,Rammelmuller:2018hnk}. A prototypical example of the former is the Hubbard model, which has attracted a great deal of attention due to its simplicity coupled with the fact that it can capture the physics of high-temperature superconductors~\cite{Baeriswyl,Lee:2006zzc}. The Hubbard model on any bipartite lattice at half-filling is free from the sign problem, but as soon as non-zero chemical potential or frustration appear in the Hamiltonian, the sign problem emerges.  One could actually claim that systems plagued by the sign problem are not the exception but, rather, constitute the majority of interesting systems. Troyer and Wiese~\cite{Troyer:2004ge} have shown that the sign problem is an NP-complete problem in one particular system, and thus, a generic solution to all sign problems is highly unlikely to be found. A brute force approach to deal with a complex action is to separate the imaginary part of the action and absorb it in the observable.
This method, known as reweighting, is based on the identity
\begin{align}
\label{eq:reweighting_identity}
    &\langle\mathcal{O}\rangle = \frac{1}{\mathcal{Z}} \int\mathcal{D}\Phi\,\mathcal{O}[\Phi]\,e^{-S[\Phi]} 
    =\frac{\langle\mathcal{O}e^{-iS_I}\rangle_{S_R}}{\langle e^{-iS_I}\rangle_{S_R}},&
\end{align}
where $S=S_R + i S_I$, and the angular brackets denote an ensemble average with a measure  $\mathcal{D}\Phi\;e^{-S_R}$.
Despite this being a mere rewriting, its practical evaluation is exponentially difficult due to the sign problem.
The technical issue is the overlap of the ensemble sampled employing only the real part of the action, with the original target ensemble that involves the full action.
The quantity $\langle e^{-S_I}\rangle_{S_R}$ can be understood as a ratio of two partition functions
$\nicefrac{\mathcal{Z}}{\mathcal{Z}_{\mathrm{pq}}} = e^{-V \beta \Delta f}$, where $\mathcal{Z}_{\mathrm{pq}}=\int\mathcal{D}\Phi\,e^{-S_R[\Phi]}$ is the phase quenched partition function. 
We have introduced the volume $V$, inverse temperature $\beta$, and  $\Delta f$, which is the free energy density difference between the two ensembles. The connection between $V$, $\beta$ and the average sign  shows that the severity of the sign problem scales exponentially with the volume and the inverse temperature.

Recently, a lot of progress has been made by exploring the idea of permitting the field variables to assume a complex value whereby it has been demonstrated for some systems that the severe sign problem of the original formulation is alleviated or sometimes even eliminated. This idea, which can easily be demonstrated in simple, one-dimensional integrals, has found nice applications in several non-trivial physical systems. We concentrate on the method of Lefschetz  thimbles~\cite{Witten:2010zr,Witten:2010cx,Cristoforetti:2012su,Cristoforetti:2013qaa,Cristoforetti:2013wha,Fujii:2013sra,Fujii:2015bua,Tanizaki:2015rda,Kanazawa:2014qma,Alexandru:2015xva,Alexandru:2015sua,Cristoforetti:2014gsa,DiRenzo:2015foa,Alexandru:2016ejd,DiRenzo:2017igr,Alexandru:2018brw,Alexandru:2018ngw,Alexandru:2018ddf,Bluecher:2018sgj}, aiming to demonstrate its potential applicability to the Hubbard model away from half-filling. We start with a short introduction to the formalism, and proceed with the general study of the saddle points, which is an essential ingredient of the method. Finally, we give several examples of Monte Carlo calculations over manifolds in complex space showing that the sign problem can be substantially weakened.  

{\it Lefschetz thimbles formalism.}
Let us consider the continuation of the functional integral~\eqref{eq:reweighting_identity} into the domain of complex-valued  fields, $\Phi \in \mathbb{C}^N$. Due to Cauchy's theorem, one can choose any appropriate contour in complex space as long as the integral still converges at infinity and no poles of the integrand are crossed during the shift of the contour. $e^{-S[\Phi]}$ doesn't have poles in our case and the convergence of integrals is guaranteed by the choice of the integration contour, as shown below.  We now define a particularly useful representation, which is known as the Lefschetz thimble decomposition of the integral \cite{Witten:2010zr,Witten:2010cx},
\begin{eqnarray}
\label{eq:thimbles_sum_and_integral}
\mathcal{Z}\! =\! \int_{\mathbb{R}^N}\! \mathcal{D} \Phi e^{-S[\Phi]}\! =\! \sum_\sigma k_\sigma \mathcal{Z_\sigma},\quad
\mathcal{Z_\sigma} \! =\!\int_{\mathcal{I}_\sigma} \! \mathcal{D} \Phi e^{-S[\Phi]},
\end{eqnarray}
where $\sigma$ labels all complex saddle points $z_\sigma \in \mathbb{C}^N$ of the action, which are determined by ${\left.{\partial S/\partial \Phi}\right| }_{\Phi=z_\sigma} = 0$.

The integer-valued coefficients $k_\sigma$, are the intersection numbers and $\mathcal{I_\sigma}$ are the Lefschetz thimble manifolds attached to the saddle points $z_\sigma$. The relation~(\ref{eq:thimbles_sum_and_integral}) holds if the saddle points are non-degenerate and isolated (for a generalization to the case of gauge theory see \cite{Witten:2010cx}). 

The Lefschetz thimble manifold is the union of all solutions of the gradient flow (GF) equations
\begin{equation}
\label{eq:flow}
\frac{d\Phi}{d\tau}=\overline{ \frac{\partial S}{\partial \Phi}},
\end{equation}
which start from the corresponding saddle point: $~\Phi\in\mathcal{I}_\sigma: \Phi(\tau\rightarrow -\infty) \rightarrow z_\sigma$. The intersection number, $k_\sigma = \langle \mathcal{K}_\sigma, \mathbb{R}^N \rangle$, is determined by counting the number of intersections of the so-called anti-thimble, $\mathcal{K}_\sigma$, with the original integration domain, $\mathbb{R}^N$. The anti-thimble $\mathcal{K}_\sigma$ consists of all possible solutions of the GF equation~(\ref{eq:flow}) which end up at a given saddle point $z_\sigma$:  $\Phi\in\mathcal{K}_\sigma:\Phi(\tau)=\Phi, \Phi(\tau\rightarrow + \infty) \rightarrow z_\sigma$.

Both thimbles and anti-thimbles are real, $N$-dimensional manifolds embedded in $\mathbb{C}^N$. Two key properties of thimbles are that the real part of the action, $\mbox{Re}\, S$, monotonically increases along it starting from the saddle point and that the imaginary part of the action, $\mbox{Im}\, S$, stays constant along it. The first property guarantees the convergence of individual integrals in~(\ref{eq:thimbles_sum_and_integral}), while the second one obviously makes the method attractive for attempts to weaken the sign problem.

\begin{figure}
        \centering
        \includegraphics[scale=0.19, angle=0]{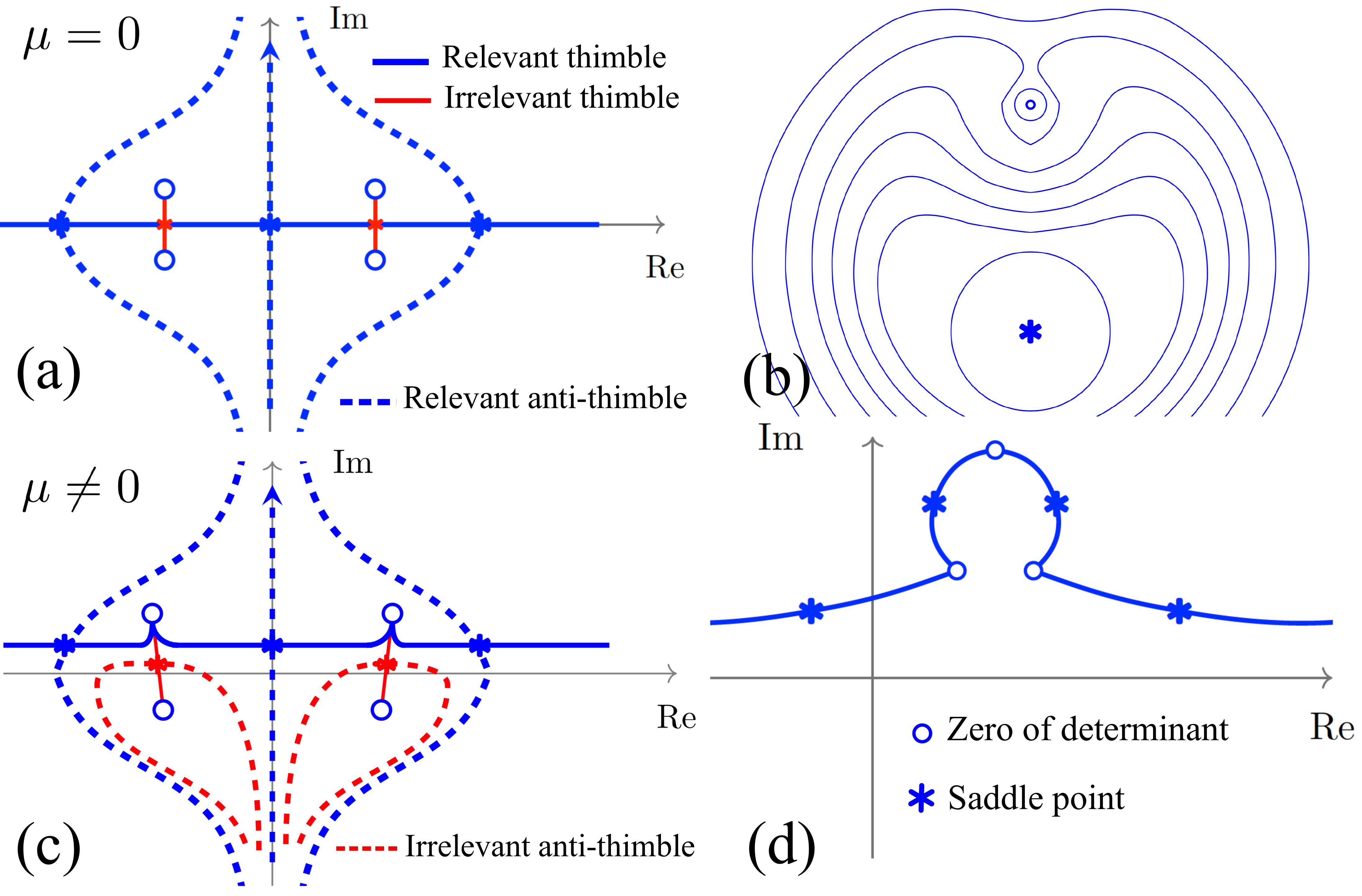}
        \caption{(a) Stokes phenomenon at half filling (1D). (b) The contour line plot for the action in 2D, which hosts relevant saddle, zero of determinant and irrelevant saddle (upper part of the figure). (c) Shift of relevant and irrelevant saddles into complex plane (1D). (d) ``Vertically'' oriented thimbles.}
        \label{fig:schemes_thimbles}
\end{figure}

Finally,~(\ref{eq:thimbles_sum_and_integral}) can be written as,
\begin{eqnarray}
\label{eq:thimbles_sum_with_phases}
\mathcal{Z} = \sum_\sigma k_\sigma e^{-i\, \mathrm{Im}\, S(z_\sigma)} \int_{\mathcal{I}_\sigma} \mathcal{D} \Phi\, e^{-\mathrm{Re} \, S(\Phi)}.
\end{eqnarray}
 
A thimble is classified as ``relevant'' if it has a nonzero intersection number, $k_\sigma$, and thus participates in the sum. There is one peculiarity associated with the so-called Stokes phenomenon, when several saddles are connected by one thimble.
Let us consider the properties of saddles points if the Stokes phenomenon occurred at half-filling, when there is no sign problem and all relevant thimbles and saddle points are confined within $\mathbb{R}^N$. In this case, all eigenvectors of the Hessian matrices $\Gamma_\sigma$ for saddles located within $\mathbb{R}^N$ have their components either purely real or imaginary. At the local minima of the action within $\mathbb{R}^N$, which is a relevant saddle point, all real eigenvectors of $\Gamma_\sigma$ correspond to positive eigenvalues. However, it can happen that some $\mathcal{N}_\sigma>0$ real eigenvectors correspond to negative eigenvalues of $\Gamma_\sigma$. This situation with two local maxima between three local minima of the action is illustrated in Fig.~\ref{fig:schemes_thimbles}\textcolor{red}{a}. Because the thimbles attached to local minima cover the entire $\mathbb{R}^N$, saddles which have at least one real eigenvector corresponding to a negative eigenvalue of $\Gamma_\sigma$, do not participate in the sum~(\ref{eq:thimbles_sum_with_phases}), and thus are irrelevant. A two-dimensional example with  $\mathcal{N}_\sigma=1$ (Fig.~\ref{fig:schemes_thimbles}\textcolor{red}{b}) shows that one actually does not need to connect two relevant saddles for this, as the irrelevant saddle can be fully embedded into one thimble. Simply counting the intersection points is impossible in this case as $\dim (\mathbb{R}^N \cap \mathcal{K}_\sigma )=\mathcal{N}_\sigma>0$ for such saddles.

The initial sign problem is now split into two parts. One part comes from the constant phase factors $e^{-i\, \mathrm{Im}\, S(z_\sigma)}$, and the second comes from fluctuations of the complex measure $\mathcal{D} \Phi$ in the integration over the thimble. Both of these issues will be addressed in our study. We will start from the search for saddle points and then we will give an estimate for the fluctuations of the complex measure.

{\it The model.}
We consider the Hubbard model on the hexagonal lattice at finite chemical potential. We start from the Hamiltonian written in the particle-hole basis in order to have a manifestly positive weight for the auxiliary field configurations at half filling
\begin{eqnarray}
  \label{eq:Hamiltonian_el_hol}
  \hat{\mathcal{H}}\! =\! -\kappa\! \sum_{\langle x,y\rangle} (  \hat a^\dag_{x} \hat a_{y} \!+\! \hat b^\dag_{x} \hat b_{y} \!+\! \mbox{h.c} ) \!+\! \frac{U}{2} \sum_{x} \hat q_x^2 + \mu  \sum_x \hat q_x,
\end{eqnarray}
where $\hat a^\dag_{x}$ and  $\hat b^\dag_{x}$ are creation operators for electrons and holes,  $\hat q_x=\hat n_{x, el.} - \hat n_{x, h.}={\hat a^\dag_{x}} \hat a_{x}- {\hat b^\dag_{x}} \hat b_{x}$ is the charge operator, $\kappa$ is the hopping parameter, $U$ is the Hubbard interaction, and $\mu$ is the chemical potential. Due to the van Hove singularity in the density of states, one can identify a clear scale, where new physics is expected at $\mu=\kappa$. Special attention will be paid to this value of $\mu$ in our study.

One can obtain an additional, nonphysical, degree of freedom in the Hamiltonian, by splitting the interaction term in the following way
\begin{eqnarray}
\frac{U}{2}\hat q_x^2 = \frac {\alpha U}{2}\hat q_x^2 - \frac{(1-\alpha) U}{2} \hat s_x^2 + (1-\alpha) U \hat s_x,
\label{eq:split_int}
\end{eqnarray}
where $\hat s_x = \hat n_{x, el.} + \hat n_{x, h.}$. Now, we can introduce two continuous auxiliary fields simultaneously by applying the standard Hubbard-Stratonovich (HS) transformation to each four-fermion term in~(\ref{eq:split_int}). The parameter $\alpha \in [0,1]$, defines the balance between auxiliary fields coupled to charge ($\hat q_x$) and spin ($\hat s_x$) density. Extreme cases $\alpha=0$ and $\alpha=1.0$ mean that only one auxiliary field is left. This decomposition of the interaction term is not the most general, but it is commonly used in QMC algorithms with continuous auxiliary fields \cite{complex1,Assaad_complex}.

\begin{figure}
        \centering
        \includegraphics[scale=0.26, angle=0]{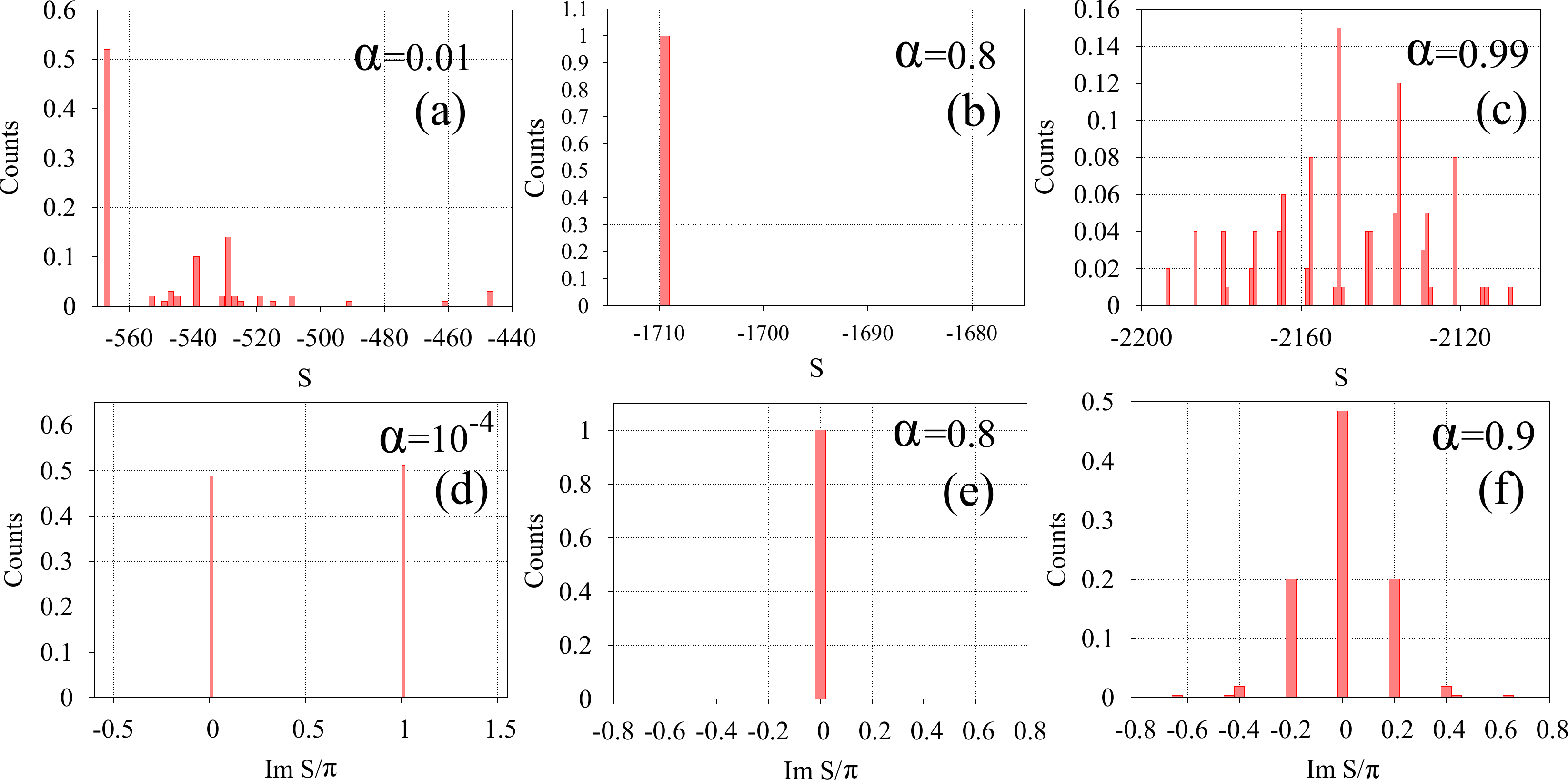}
        \caption{The distribution of the action at the saddle points for a $6\times6$ lattice with $N_t=256$ and $\beta=20.0$ at the interaction strength corresponding to the phase transition from SM (semimetal) to AFM (antiferromagnetic) phase: $U=3.8 \kappa$. The upper row (a-c) shows the real part of the action at half-filling.
        The lower row (d-f) shows the distribution of the imaginary part of the action for saddle points when $\mu=\kappa$. Three different values of $\alpha$ (see eq. (\ref{eq:split_int}))  are considered in both cases.}
        \label{fig:alpha_histogram}
\end{figure}

Further construction of the path integral is straightforward and the detailed derivations can be found in \cite{ITEPRealistic, SmithVonSmekal,Assaad_complex}. Here we simply state the explicit form which we have used in our calculations,
\begin{align}
  &\mathcal{Z}_c = \int \mathcal{D} \phi_{x,t} \mathcal{D} \chi_{x,t} e^ {-S_{\alpha}}  \det M_{el.} \det M_{h.},&  \\
   &S_\alpha[\phi_{x,t},\chi_{x,t}]\!=\!\sum_{x,t}  \frac  {\phi_{x,t}^2} {2 \alpha \delta U}  + \sum_{x,t}  \frac {(\chi_{x,t} + (1-\alpha) \delta U)^2} {2 (1-\alpha) \delta U},&\nonumber
  \label{eq:Z_continuous}
\end{align}
where the fermionic operators are given by
\begin{eqnarray}
 M_{el.,h.} = I +\prod^{N_{t}}_{t=1} \left({ e^{-\delta \left(h\pm\mu\right)} \diag{ e^{\mp i \phi_{x,t}+\chi_{x,t}} } }\right), 
  \label{eq:M_continuous}
\end{eqnarray}
where $h$ is the matrix of one-particle tight-binding Hamiltonian in~(\ref{eq:Hamiltonian_el_hol}). The field $\phi_{x,t}$ is coupled to charge density, and the field $\chi_{x,t}$ to spin density. The full action involves both the bosonic action as well as the logarithm of the fermionic determinants,  $S=S_{\alpha} - \ln (\det M_{el.} \det M_{h.})$. The total number of the fields is $2 N_s N_t$, where $N_s$ is the spatial size and $N_t$ is the Euclidean time extent of the lattice.

\begin{figure}
        \centering
        \includegraphics[scale=0.085, angle=0]{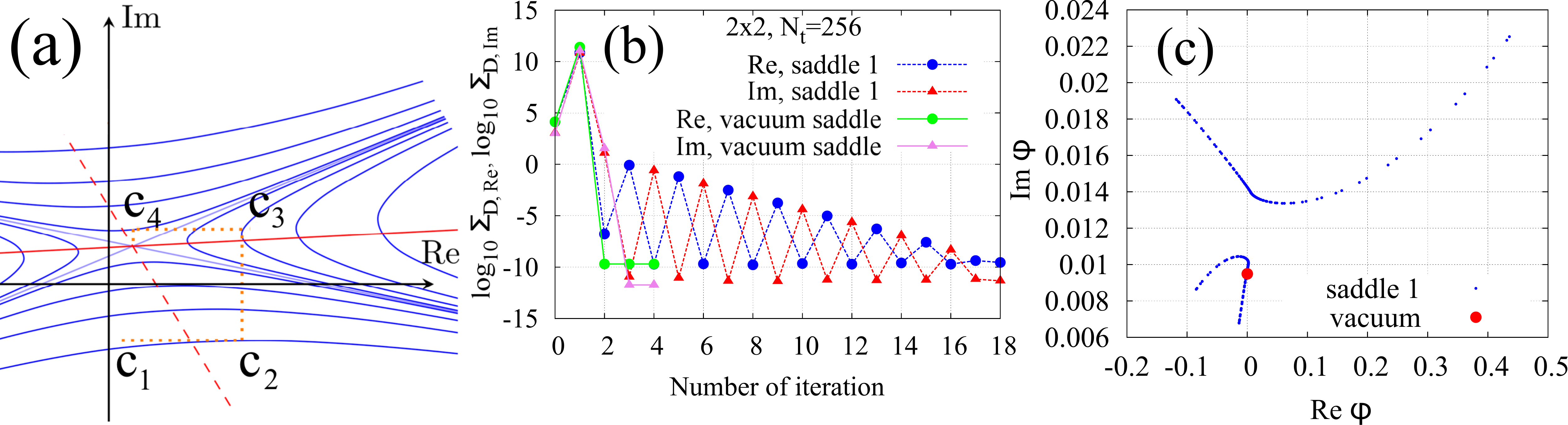}
        \caption{(a) Schematic illustration of the search algorithm for complex saddle points (1D case): $c_1$ is the initial point; $c_2$, $c_3$, $c_4$ are the final points of the corresponding iteration of the GF. 
        (b) Example of a search processes for a $2\times2$ lattice with $N_t=256$, $\beta=20.0$, $U=2.0 \kappa$, $\mu=\kappa$ and $\alpha=1.0$. The shorter process converges to the vacuum saddle point and the longer one shows convergence to a non-vacuum, localized saddle point. These saddles are shown in the plot (c). $\chi_{x,t}=0$, complex values of all $\phi$-fields are projected onto a single complex plane.}
        \label{fig:general_complex_search }
\end{figure}

{\it Saddle points study.} We begin with a study of the saddles at half-filling. Unlike \cite{Ulybyshev:2017}, where analytic results could be obtained for very small systems, we devise a hybrid approach. First, we generate lattice configurations using hybrid Monte Carlo (HMC) with exact fermionic forces \cite{PhysRevB.99.205434}. Second, we numerically integrate the GF equations starting from these configurations for a large enough flow time to reach local minima of the action. The distribution of lattice ensembles, taken after employing the GF procedure, gives an accurate characterization of the relevant saddle points at half-filling. Results are shown in Figs.~\ref{fig:alpha_histogram}\textcolor{red}{a-c}. At half-filling we can not work exactly at $\alpha=0$ and $\alpha=1.0$ because HMC is not ergodic there \cite{PhysRevB.98.235129}, thus $\alpha=0.01;0.8;0.99$. At small $\alpha$, $\chi_{x,t}$ dominates in saddles,  while $\phi_{x,t}=0$.  The lowest bar in Fig.~\ref{fig:alpha_histogram}\textcolor{red}{a} corresponds to two identical mean-field saddle points, which describe sublattice magnetization ($\phi_{x,t}=0$ and $\chi_{x,t}=\pm m$, depending on sublattice). Other saddles  correspond to various single- and multi-instanton solutions. At large $\alpha$ (Fig.~\ref{fig:alpha_histogram}\textcolor{red}{c}), $\phi_{x,t}$ is dominant,  while $\chi_{x,t}=0$. The lowest bar is the vacuum saddle ($\phi_{x,t}=\chi_{x,t}=0$). The next bar contains equal contributions from two localized field configurations (so-called ``blob'' and ``anti-blob''). All other saddles are various combinations of several blobs and anti-blobs. Finally, there is intermediate regime at $\alpha\approx0.8$ (Fig.~\ref{fig:alpha_histogram}\textcolor{red}{b}), where only vacuum saddle was found.  More detailed description of saddles including results for larger lattices can be found in \cite{OurThimblesPRB:2019}.

 \begin{figure}
        \centering
        \includegraphics[scale=0.18, angle=0]{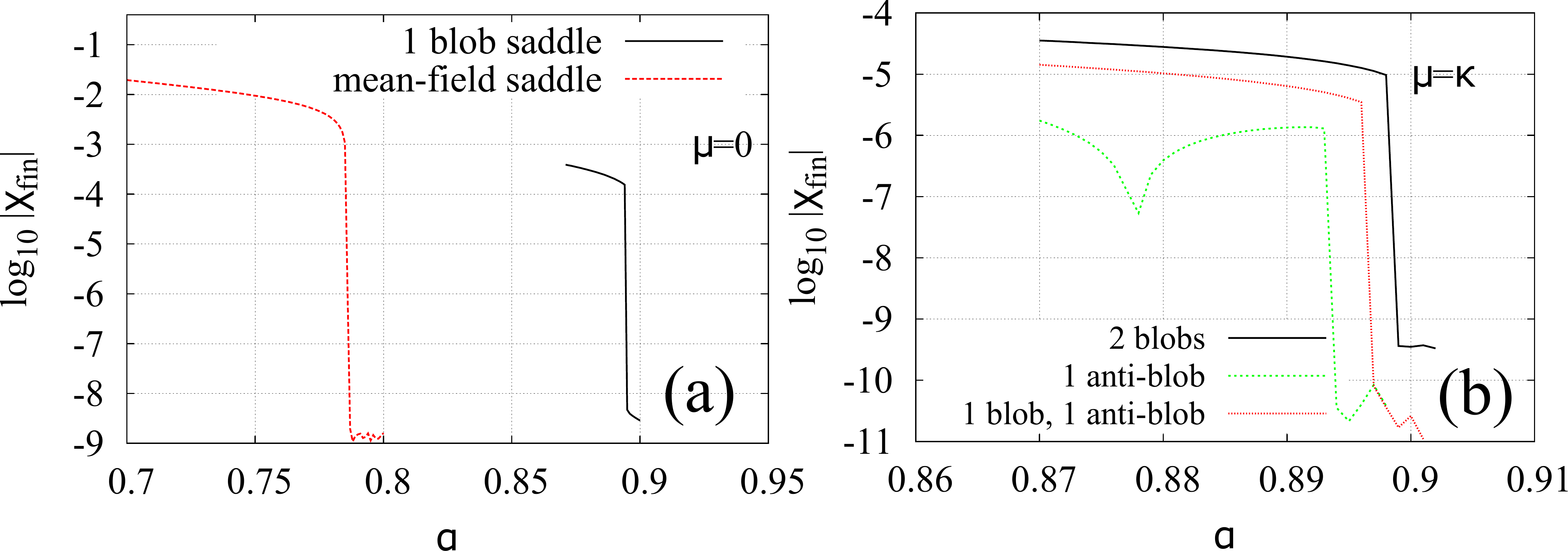}
        \caption{Results of the study of the $\alpha$-dependence of saddle points at half-filling (a) and at $\mu=\kappa$ (b). In both plots, we depict the value of $\chi_{x,t}$ at $x=0$, $t=0$ in the end of the GF, started from slightly disturbed saddle. All calculations are done for a $6\times6$ lattice with $N_t=256$ and $\beta=20.0$, $U=3.8 \kappa$.}
        \label{fig:history_alpha}
\end{figure}

Another procedure should be used away from half-filling. Downward GF follows into saddle point only if initial configuration was exactly on thimble. At $\mu \neq 0$ it is possible only at $\alpha=0.0$, but HMC is again not ergodic there \cite{Assaad_complex}. Thus we generate initial configurations using phase quenched HMC along some contour in $\mathbb{C}^N$, uniformly shifted from $\mathbb{R}^N$, in order to approach the thimble (e.g. using the notion about the shift of vacuum saddle into complex space at $\mu \neq 0.0$). Subsequently, we launch the procedure, illustrated in Fig.~\ref{fig:general_complex_search }\textcolor{red}{a}, for a single complex field. The minimization procedure proceeds by alternation of GF for constant imaginary and real parts of the field. Even iterations are GF in downward direction with fixed $\mbox{Im} \Phi_j=\Phi^{(R)}_j$, where $\Phi_j \equiv \Phi^{(R)}_j + i\Phi^{(I)}_j$ represents both complex auxiliary fields. The flow stops when it reaches the local minimum. Odd iterations are upward GF with fixed $\mbox{Im} \Phi_j=\Phi^{(I)}_j$. This flow stops when it reaches a local maximum or zero of determinant, where $\mbox{Re} S \rightarrow \infty$. Convergence can be controlled by monitoring $\Sigma_{D, Re/Im} \equiv \sum_i | \partial \mbox{Re} S/ \partial \Phi^{(R/I)}_i |$ after each iteration, with $\Sigma_{D, Re}$ reaching the level of numerical errors (typically $10^{-10}$) at even iterations and $\Sigma_{D, Im}$ at odd ones (if the flow didn't collide with a zero of determinant). Examples are show in Fig.~\ref{fig:general_complex_search }\textcolor{red}{b}.

\begin{figure}
        \centering
        \includegraphics[scale=0.3, angle=0]{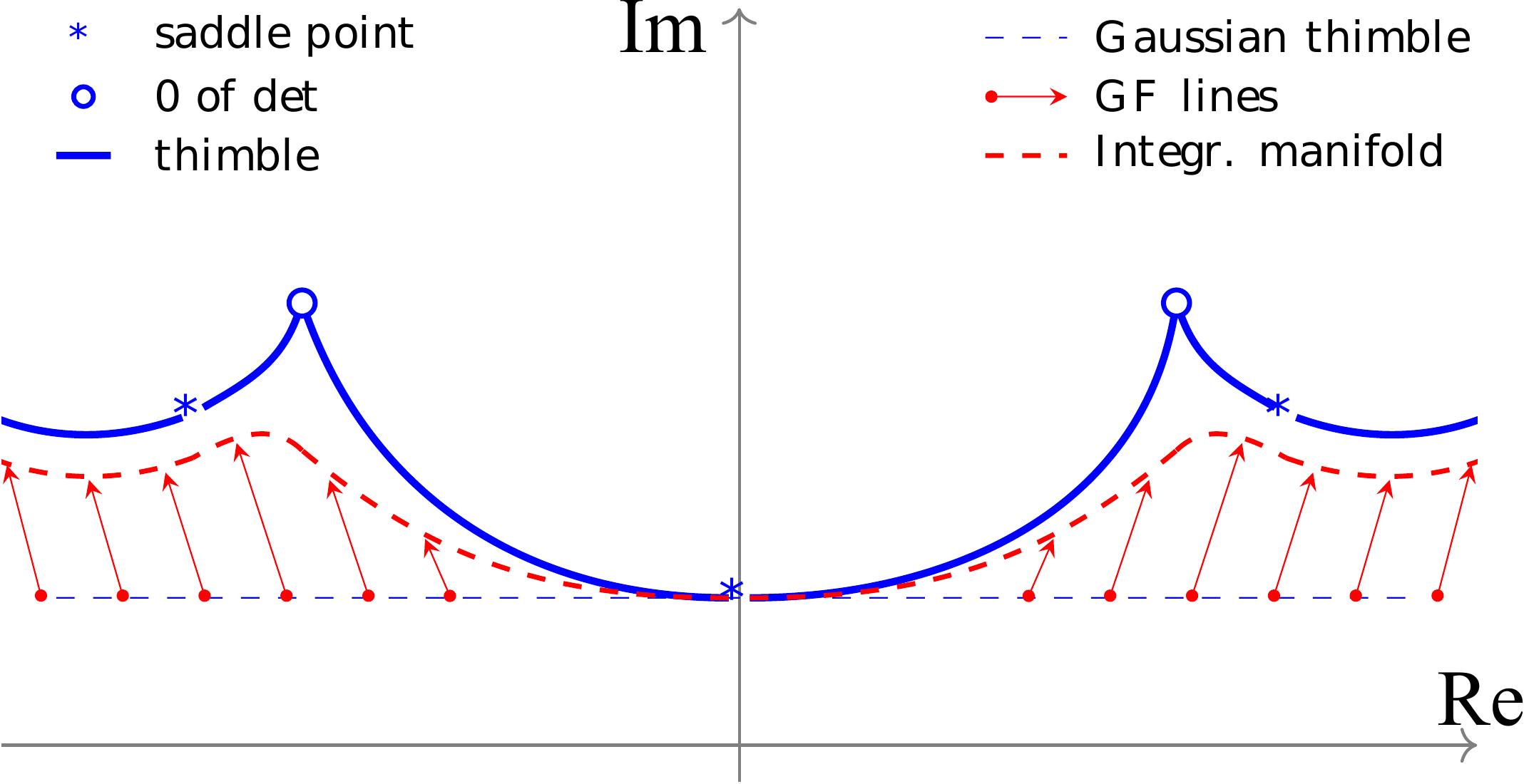}
        \caption{Schematic illustration of HMC with gradient flow.}
        \label{fig:schemes_hmc_flow}
\end{figure}

This procedure does not converge for all saddles. The criterion for the convergence of the procedure can be understood in terms of the Hessian matrix, which is defined as, 
$\Gamma = \begin{pmatrix} 
A & C \\
C & B 
\end{pmatrix}$.
We express the Hessian in terms of $2N^2_s N_{\tau} \times 2N^2_s N_{\tau}$ blocks $A_{i,j} \equiv \partial^2 \mbox{Re} S/\partial \Phi^{(R)}_i \partial \Phi^{(R)}_j$, $B_{i,j} \equiv \partial^2 \mbox{Re} S/\partial \Phi^{(I)}_i \partial \Phi^{(I)}_j$, and $C_{i,j} \equiv \partial^2 \mbox{Re} S/\partial \Phi^{(R)}_i \partial \Phi^{(I)}_j$. Using these matrices, the minimization procedure is guaranteed to converge if both $A$ and $-B$ are positive-definite, and each of the eigenvalues, $\lambda_i$, of the matrix $A^{-1} C B^{-1} C$, which characterizes the update of the fields after each iteration, satisfy $|\lambda_i| < 1$. The latter is actually a restraint for $|\arg \partial_i \partial_j S|$. If all of the second derivatives are real, $C=0$, and thus $|\lambda_i| = 0$. If $|\arg \partial_i \partial_j S|$ increases, with $A$ and $(-B)$ still remaining positive-definite, the thimble in the vicinity of the saddle point is no longer parallel to $\mathbb{R}^N$, but starts to ``rotate'' in complex space. In the $1$D case illustrated in Fig. ~\ref{fig:general_complex_search }\textcolor{red}{a}, $|\lambda| < 1$ simply means $|\arg \partial^2 S |_{z_\sigma}|<\pi/4$.

The distribution of $\mbox{Im} S$ over saddle points obtained using this method is shown in Fig.~\ref{fig:alpha_histogram}\textcolor{red}{d-f}. At $\alpha= 0.8$, we detect again only the vacuum saddle, which is uniformly shifted into complex space, $\phi_{x,t}= i \phi_0$, $\chi_{x,t}=0$. However, we should stress that unlike $\mu=0$ case the distribution is only approximate, since the initial configurations for the iterations are not on thimble. Furthermore, ''vertically oriented'' saddles (see Fig.~\ref{fig:schemes_thimbles}\textcolor{red}{d} for 1D example) can be missed due to limitations in the convergence of the algorithm as previously described.

Fig.~\ref{fig:history_alpha} provides some insight into the nature of the optimal regime around $\alpha=0.8$. Starting from half-filling in Fig.~\ref{fig:history_alpha}\textcolor{red}{a}, we see the lower boundary of this region in $\alpha$ which was identified by the splitting of the vacuum saddle into two mean-field saddles. This splitting is observed by launching GF from a slightly perturbed vacuum (Gaussian noise added in $\phi_{x,t}$ and $\chi_{x,t}$). If $\chi_{x,t}$ returns to zero, the vacuum is stable, otherwise the final value of $\chi_{x,t}$ is non-zero. The upper boundary is more difficult to determine. We use the symmetry, $S(\phi_{x,t}, \chi_{x,t}) = S(\phi_{x,t}, - \chi_{x,t})$, and the fact that the saddles are located at $\chi_{x,t}=0$ for large $\alpha$. It follows that the Hessian matrix is block-diagonal: $\partial^2 S / \partial \chi_{x,t} \partial \phi_{x,t}=0$. According to what was previously stated about the saddles at half-filling, we can study the relevance of saddles separately for $\phi$- and $\chi$-directions, as it suffices to find at least one instability (negative eigenvalue of $\Gamma$, with real eigenvector). We first study the $\phi$-direction using GF restricted to the $\phi$ fields and starting from slightly perturbed saddles. No instability was found, and the non-trivial saddles can be found for all values of $\alpha$. Finally, we use these saddles, add noise to the $\chi$ fields, and launch the restricted GF for these fields. At $\alpha \approx 0.89$, the final value of $\chi$ jumps upwards. It signals am instability in the $\chi$-channel, and thus such saddles become irrelevant. It is sufficient to study only one-blob configurations, as being the building block for all other saddles, we expect other saddles to behave similarly \cite{OurThimblesPRB:2019}. 

Similar calculations were made for $\mu=\kappa$, where we have used GF restricted to $\mbox{Re} \chi$ (Fig.~\ref{fig:history_alpha}\textcolor{red}{b}), and essentially the same behavior was observed. This suggests that at $\alpha>0.89$, the non-vacuum saddles shift into complex space with increasing $\mu$, remaining relevant, while at $\alpha<0.89$ they begin from irrelevant ones at $\mu=0$ and proceed in the same status into complex space for $\mu\neq0$, remaining irrelevant. Another possibility is that the saddles acquire a more ``vertical'' orientation with decreasing $\alpha$ (as in Fig.~\ref{fig:schemes_thimbles}\textcolor{red}{d}). GF along $\mbox{Re} \chi$ can go away from zero in this case too.  However, there are additional arguments against this from the results of QMC simulations at different $\alpha$, described in the next section.

\begin{table}[t]
  \begin{tabular}{ | c | c | c |}
    \hline
                      & $\langle K \rangle$ & $\langle S^{(1)}_x S^{(1)}_y \rangle$   \\  \hline  \hline
    ED                     & 19.5781             & -0.14624              \\ \hline
    BSS-QMC          & 19.587$\pm$0.002    & -0.1466$\pm$0.0008    \\ \hline
    HMC-GF, $\alpha=1.0$ & 19.65$\pm$0.31      & -0.112$\pm$0.0069    \\ \hline
    HMC-GF, $\alpha=0.8$ & 19.52$\pm$0.17      & -0.142$\pm$0.0062     \\ \hline
    \hline
  \end{tabular}
\caption{Comparison of observables for exact diagonalization, BSS-QMC, and two variants of HMC-GF. Parameters of simulations: $2 \times 2$ lattice, $N_t=256$, $U=2 \kappa$, $\beta=20$, $\mu=\kappa$. }
  \label{tab:observables}
\end{table}

\begin{table}[t]
  \begin{tabular}{ |  c | c | c | c |}
    \hline
                      & $\langle \cos \mbox{Im} S \rangle$  & $\langle \cos \mbox{Arg} J \rangle$ & $\langle \Sigma_G \rangle$  \\  \hline  \hline
    BSS-QMC          & 0.2363$\pm$0.0032 &                  & 0.2363$\pm$0.0032  \\ \hline
    HMC-GF, $\!\alpha\!\!=\!\!1.0$ & 0.9627$\pm$0.0038 & 0.427$\pm$0.014  & 0.351$\pm$0.015    \\ \hline
    HMC-GF, $\!\alpha\!\!=\!\!0.8$ & 0.797$\pm$0.022   & 0.915$\pm$0.008  & 0.644$\pm$0.028    \\ \hline
    \hline
  \end{tabular}
\caption{Comparison of the sign problem for BSS-QMC (ALF) and two variants of HMC-GF. Parameters of simulations: $2 \times 2$ lattice, $N_t=256$, $U=2 \kappa$, $\beta=20$, $\mu=\kappa$.}
  \label{tab:sign}
\end{table}

{\it HMC with gradient flow.} The general scheme for the deformation of the integration contour is shown in Fig.~\ref{fig:schemes_hmc_flow}.  Following \cite{Alexandru:2016ejd}, the sequence of deformations can be summarized by
\begin{equation}
\mathcal{Z}=\int_{\mathbb{R}^N} \mathcal{D} \Phi e^{-S[\Phi + i \Phi_0]} = \int_{\mathbb{R}^N} \mathcal{D} \Phi e^{-S[\tilde \Phi]} \det J.
\label{eq:HMC_flow}
\end{equation}
First, we perform a uniform shift into the complex plane, $\Phi \rightarrow \Phi+i\Phi_0$. We work at large $\alpha$, and this shift corresponds to the thimble attached to the vacuum saddle (see Fig.~\ref{fig:general_complex_search }\textcolor{red}{c}) in the Gaussian approximation. A further shift is made using the GF equations. We denote $\tilde \Phi \in \mathbb{C}^N$ as the result of the evolution of the field determined by (\ref{eq:flow}, starting from the Gaussian thimble $\Phi + i \, \Phi_0, \Phi \in \mathbb{R}$ with flow time $\mathcal{T}$. The complex-valued Jacobian of the transformation, $J=\mathcal{D} \tilde \Phi / \mathcal{D} \Phi$, appears in the second stage. The flow time defines how close we can approach the thimble, and thus it regulates the fluctuations of $\mbox{Im} S$. The Jacobian can also contribute to the sign problem, especially in the case of ``vertically'' oriented thimbles, as shown in Fig.~\ref{fig:schemes_thimbles}\textcolor{red}{d}. 

We sample the $\Phi$ fields in $\mathbb{R}^N$ according to the distribution $e^{-S[\tilde \Phi[\Phi + i \Phi_0]]}$, using HMC, while $\tilde \Phi$ is then reconstructed with GF. The details of the algorithms can be found in \cite{OurThimblesPRB:2019}, and we refer to the algorithm as HMC-GF. The key point is that we compute the exact derivatives, $\partial \ln \det M_{el.,h.} / \partial \Phi_j$, with a Schur complement solver \cite{PhysRevB.99.205434, ULYBYSHEV2019118}. This allows us to solve the GF equations with high precision and performance, which is necessary, as GF is the basic building block of the algorithm. The dominant term in the scaling of the computational cost of the method is $N_s ^4 N_t ^2$.

\begin{figure}
        \centering
        \includegraphics[scale=0.18, angle=0]{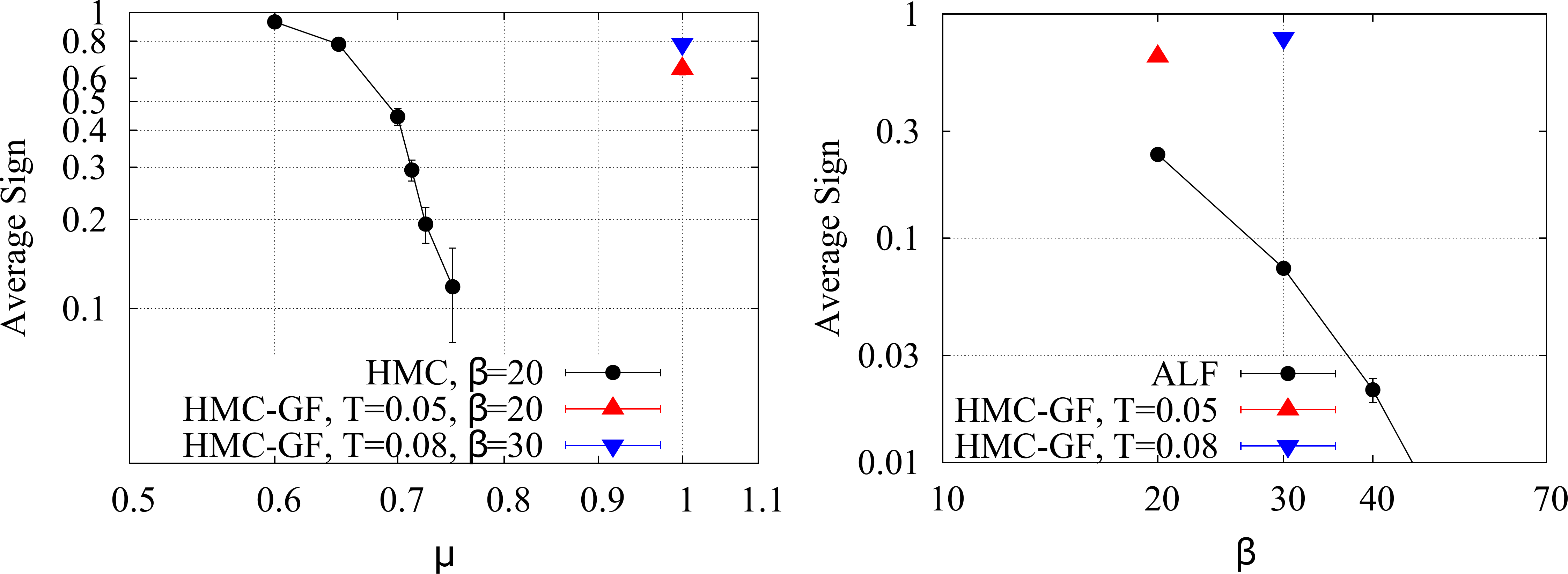}
        \caption{(a) Comparison of the sign problem in conventional HMC with real Hubbard fields and in HMC-GF. (b) Comparison of the sign problem in BSS-QMC and in HMC-GF, depending on temperature at $\mu=\kappa$. Results are shown for a $2\times2$ lattice with $U=2.0 \kappa$, $N_t=256$. $\alpha=0.8$ for all HMC-GF points.}
        \label{fig:sign_study}
\end{figure}

The Jacobian is left for the final reweighting,
\begin{equation}
\label{eq:fin_observable}
\langle\mathcal{O}\rangle =\frac{\langle\mathcal{O} e^{i \, \mbox{\small{Im}} (-S +  \ln \det J) +   \mbox{\small{Re}} (\ln \det J) } \rangle }{ \langle  e^{i \, \mbox{\small{Im}} (-S +  \ln \det J) +   \mbox{\small{Re}} (\ln \det J) }\rangle}, 
\end{equation}
where the residual fluctuations of $\mbox{Im} S$ are also taken into account.
The brackets $\langle \rangle$ denote the averaging over configurations generated with HMC-CG. Since $S\rightarrow\bar S$ if $\mbox{Re} \Phi_j$ changes sign, we use the following metrics to estimate the severity of the sign problem: $\langle \cos(\mbox{Im} S) \rangle$ and $\langle \cos(\mbox{Im}\ln \det J)\rangle$ for configurations and the Jacobian respectively, and the joint sign $\Sigma_G = \langle \cos(\mbox{Im} (-S + \ln\det J))\rangle$. We also estimate the strength of the fluctuations of the Jacobian by computing $D_J$, the dispersion of $\mbox{Re} (\ln \det J)$. 

We made the following choice for the parameters of the simulations: $2 \times 2$ lattice ($N_s=8$), $N_t=256$, $U=2 \kappa$, $\mu=\kappa$, $\beta=20$. This lattice is small enough to make the fast comparison with finite-temperature ED possible, but large enough to host non-trivial saddle points at large $\alpha$ (see Fig.~\ref{fig:general_complex_search }\textcolor{red}{c}). These saddles also decay in the $\chi$ channel at $\alpha=0.8$, similar to the $6 \times 6$ lattices studied above. $N_t$ is large enough to probe the low-temperature limit and continuous limit in Euclidean time. Also, the state-of-the-art QMC algorithm for condensed matter systems, BSS (Blankenbecler, Scalapino and Sugar)-QMC taken from the ALF package \cite{ALF2017}, experiences exponential decay of the average sign at these parameters, thus the sign problem is already strong enough.

Results for observables are displayed in the table \ref{tab:observables}. We compute the kinetic energy, $\vev{\hat{K}}$, and the nearest-neighbor correlation function for the first component of spin $\vev{\hat{S}^{(1)}_x\hat{S}^{(1)}_y}$. The study of the sign problem is summarised in Tab.~\ref{tab:sign}.  Results at $\alpha=1.0$ substantially deviate from ED, while at $\alpha=0.8$ the ED results fall within errorbars of HMC-GF calculation. It means that at $\alpha=1.0$ we indeed have ergodicity issues because there are several relevant thimbles and the flow lines collide with zeros of determinant. HMC-GF can not penetrate the border between two thimbles in such situation. At $\alpha=0.8$ the ergodicity is restored. Moreover, we do not observe the growth of the fluctuations of Jacobian, which should appear if GF approaches ``vertically'' oriented thimbles (Fig.~\ref{fig:schemes_thimbles}\textcolor{red}{d}). It means that the thimbles attached to non-vacuum saddles indeed become irrelevant or they are shortcut by the integration manifold, constructed by GF. In both cases these non-vacuum saddles are effectively not important at $\alpha\approx 0.8$.

In addition to this check, we collected smaller statistics for the same lattice, but $\beta=30$ with $N_t=384$ ($\alpha=0.8$). It appears that we need to increase the flow time from $\mathcal{T}=0.05$ for $\beta=20$ to  $\mathcal{T}=0.08$ for $\beta=30$ to keep fluctuations of $\mbox{Im} S$ roughly the same. Results are shown in Fig.~\ref{fig:sign_study} with comparison to other methods.

We finally remark that the dispersion of the Jacobian is noticeable but in general does not cause large problems. We simply need larger statistics to compensate for these additional fluctuations. We have determined $D_J=1.157$ for HMC-GF with $\alpha=1.0$ and $D_J=1.011$ for HMC-GF with $\alpha=0.8$.
However, we noticed that the properties of the Jacobian become worse at $\beta=30$: $D_J=1.68$ and $\langle \cos \mbox{Arg} J \rangle = 0.823 \pm 0.018$ in this case (compare it with $\alpha=0.8$ case in Tab.~\ref{tab:sign}). Thus, at very low temperatures and, possibly, at larger system sizes, fluctuations of the Jacobian might become a problem.

{\it Conclusions.}
We demonstrated the possibility to simplify the structure of Lefschetz thimbles decomposition by manipulating the form of the HS transformation. Several examples of large-scale HMC-GF simulations are shown for the lattices with up to $2 \times 2\times 2\times384$ sites. The correctness of results is confirmed with ED. The sign problem is substantially weakened in comparison with conventional QMC, which ensures the perspectives of the Lefschetz thimbles method for the Hubbard model.

{\it Acknowledgements}

MU would like to thank Prof. F. Assaad for enlightening discussions. SZ acknowledges stimulating discussions with Prof. K. Orginos. 
MU is supported by the DFG under
grant AS120/14-1. CW is supported by the University of Kent, School of Physical Sciences. SZ acknowledges the support of the DFG Collaborative Research Centre SFB 1225 (ISOQUANT). This work was granted access to the HPC resources of CINES and IDRIS in France under the
allocation 52271 made by GENCI. We are grateful to these computing centers for their constant help. We are grateful to the UK Materials and Molecular Modelling Hub for computational resources, which is partially funded by EPSRC (EP/P020194/1). This work was also partially supported by the HPC Center of Champagne-Ardenne ROMEO.

\bibliography{thimbles}

\end{document}